\documentclass[aps, prl, twocolumn, superscriptaddress, showpacs, byrevtex, floatfix]{revtex4-1}
\usepackage{graphicx}

\begin{document}
\title{Femtosecond Population Inversion and Stimulated Emission \\
of Dense Dirac Fermions in Graphene}
\author{T. Li}
%\email[]{ioannis@iastate.edu} 
\affiliation{Ames Laboratory and
Department of Physics and Astronomy, Iowa State University, Ames,
Iowa 50011, U.S.A.}
\author{L. Luo}
\affiliation{Ames Laboratory and Department of Physics and
Astronomy, Iowa State University, Ames, Iowa 50011, U.S.A.}
\author{M. Hupalo}
%\email[]{nhshen@ameslab.gov} 
\affiliation{Ames Laboratory and
Department of Physics and Astronomy, Iowa State University, Ames,
Iowa 50011, U.S.A.}
\author{J. Zhang}
\affiliation{Ames Laboratory and Department of Physics and
Astronomy, Iowa State University, Ames, Iowa 50011, U.S.A.}
\affiliation{Department of Physics, College of William and Mary, Williamsburg, Virginia 23187}
\author{M. C. Tringides}
\affiliation{Ames Laboratory and Department of Physics and
Astronomy, Iowa State University, Ames, Iowa 50011, U.S.A.}
\author{J. Schmalian}
\affiliation{Ames Laboratory and Department of Physics and
Astronomy, Iowa State University, Ames, Iowa 50011, U.S.A.}
\affiliation{Institute for Theory of Condensed Matter and Center for Functional Nanostructures,
Karlsruhe Institute of Technology, Karlsruhe 76128, Germany}
\author{J. Wang}
\affiliation{Ames Laboratory and Department of Physics and
Astronomy, Iowa State University, Ames, Iowa 50011, U.S.A.}
\date{\today}

\begin{abstract}
We show that strongly photoexcited graphene monolayers with $35$fs pulses 
\emph{quasi-instantaneously} build up a \emph{broadband, inverted} Dirac
fermion population. Optical gain emerges and directly manifests itself via a
negative conductivity at the near-infrared region for the first 200fs, where stimulated emission
completely compensates absorption loss in the graphene layer. Our
experiment-theory comparison with two distinct electron and hole chemical
potentials reproduce absorption saturation and gain at $40$fs, revealing,
particularly, the evolution of the transient state from a hot classical gas
to a dense quantum fluid with increasing the photoexcitation.
\end{abstract}
\maketitle

Graphene is gradually emerging as a prominent platform for ultrafast
photonics and optoelectronics \cite{BonaccorsoNaturephotonics2010,RMP2009, KFM2008}. 
%Graphene is unique due to its linear Dirac spectrum $\varepsilon _{\pm
%}\left( p\right) =\pm vp$, with both vanishing gap and density of states at
%the neutrality point \cite{RMP2009}. 
%Besides the well-established distinctive transport properties, 
Growing evidence was demonstrated in, e.g., broadband transparency and
universal absorption from the near-infrared to visible\cite{NairScience2008}, carrier dynamics \cite{DawlatyAPL2009},
saturable absorption \cite{SunACSNano2010}, pulsed
photoluminescence \cite{LuiPRB2010, LuiPRL2010,StohrPRB2010}, and
coherently-driven photo-currents \cite{SunNanoLett2010}. For graphene to
play a significant role in ultrafast laser technology or telecommunications
that exceed semiconductor nanostructure performance, it is vital to
investigate femtosecond nonlinearities of strongly photoexcited states.
Prior time-resolved studies in graphene have been mostly concerned with the
weak excitation regime where the photoexcited carrier density is much
smaller than the initial background carrier density. In this case a linear
power dependence of transient signals was observed \cite{DawlatyAPL2009, BreusingPRL2009, NewsonOE2009, SunPRL2008}.  

Ultrafast photoexcitations strongly alter the thermodynamic equilibrium of
electronic states and lead to a series of temporally overlapping rapid
processes in graphene, as illustrated in Fig. 1(a). First, during or
immediately following the pulse duration $\tau _{p}$, photoexcitations are $%
coherent$. Then, electron-electron collisions lead to decoherence and
eventually to a quasi-thermal transient distribution after a time $\tau _{%
\mathrm{th}}$. Finally, for longer times via coupling to phonons, the system
relaxes back to equilibrium via cooling of the hot carriers ($\tau _{\mathrm{%
c}}$) and recombination of electron-hole pairs ($\tau _{\mathrm{r}}$). In
most semiconductors and their nanostructures, where $\tau _{\mathrm{th}}\gg
\tau _{\mathrm{p}}$ for $\sim $10 fs laser pulses \cite{oudarPRL1985},
ultrafast nonlinear photoexcitations lead to a largely non-thermal, peaked
distribution close to the pump photon energy and \textit{state
filling} dominates on the 10s of \textrm{fs} time scale [first panel, Fig.
1(b)]. 

\begin{figure}[tbp]
\includegraphics[scale=0.38]{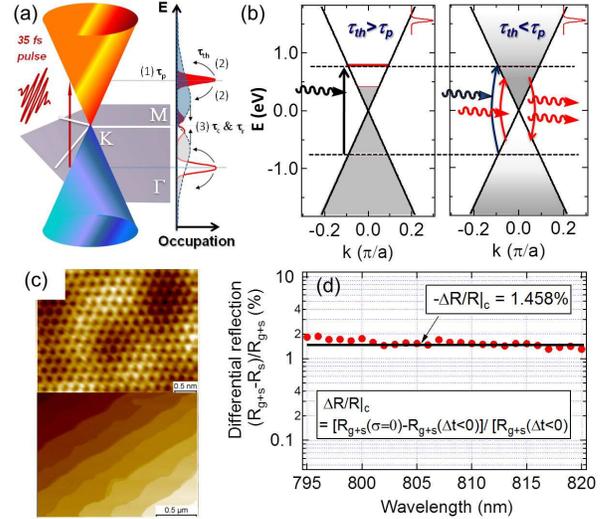}
\caption{(color online) (a) Schematics of ultrafast optical interband
excitations. (b) Dispersion of our electron-doped graphene monolayers ($%
\protect\mu =0.4eV$) illustrating state filling (left) and band filling
(right) that leads to stimulated emission from a broadband, inverted
population (red arrow). Also shown together is the pump pulse spectrum. (c)
STM images of tomography of the sample used. (d) The static differential
reflectivity spectra (red dots). The threshold $\Delta R/R_{c}$ for zero
conductivity can be directly determined $\sim$-1.46$\%$ (black line, see
text for details).}
\label{Fig1}
\end{figure}

In this letter, we demonstrate that graphene is in the opposite limit $\tau
_{\mathrm{th}}<\tau _{\mathrm{p}}$ for 35 fs pulses, where a broadband quasi-equilibrium, yet
population inverted dense Fermi system emerges during the pulse propagation.
The photoexcited phase space is quickly depleted via carrier-carrier
collisions, leading to \textit{band filling} [second panel of Fig. 1(b)].
Most intriguingly, we demonstrate that broadband gain emerges below the
excitation photon energy via stimulated emission. 
%The transient state of
%dense Dirac fermions %is stabilized by the phase space
%constraints of the relativistic spectrum, and it 
%exhibits  
The broadband optical gain directly manifests itself via a
remarkable negative conductivity at the near-infrared probe energy detuning even $\sim $%
400 meV below the excitation level within hundreds of fs. These results  can be quantitatively
described in terms of  distinct electron and hole chemical potentials, which
evolves from a hot classical gas to a dense quantum fluid with increasing
the excitation from the linear to the highly nonlinear regime. Such
femtosecond build-up of high-density and broadband population inversion has
implications in advancing gaphene-based above-terahertz speed modulators,
saturable absorbers and gain mediums.

A Ti:Sapphire amplifier with central wavelength $800\mathrm{nm}$ ($\hbar
\omega =1.55 $\textrm{eV) } and $\tau_{p} =35\mathrm{fs}$ is used to pump an
optical parametric amplifier (900-2400 nm) to produce near-infrared probes
below the 1.55 eV excitation. Ultrafast degenerate and non-degenerate
differential reflectivity changes $\Delta R/R $ with $\sim 40$ \textrm{fs}
time resolution are recorded with tunable pump fluences, which exhibits no pump polarization dependence. 
Our epitaxial
graphene monolayer sample was grown by thermal annealing in ultrahigh vacuum
on a Si-terminated 6H-SiC(0001). 
The Fermi energy is $\sim 0.4$ \textrm{eV} in the sample (Fig. 1b)
reflecting the substrate-induced electron doping as reported \cite%
{GierzNanoLett2008, Hupalo09}. Fig.1(c) shows the STM characterizations of tomography
of our samples used, which show homogenous carbon monolayer in atomic length
scales (top) and $\mu m$ scales with smooth overgrowth on the steps of SiC
surface (bottom). The strong dependence of the $6\sqrt{3}\times 6\sqrt{3}$%
R30 reconstruction modulation with the bias voltage confirm the single layer
graphene, with the homogeneity of the sample better than 90$\%$ across the
entire probe region. The monolayer thickness is further confirmed by the optical
deferential reflection spectra [Fig. 1(d)], determined by the
measurements with (R$_{g+s}$) or without (R$_s$) graphene on SiC substrate. 
% (see the next paragraph) \cite{suppl}.

To understand optical response in graphene,  we expand the solution of the
Fresnel equation with respect to $\sigma /c$ (of the order of the fine
structure constant of quantum electrodynamics $\ \alpha _{\mathrm{QED}%
}=e^{2}/\left( \hbar c\right) \approx 1/137\ll 1$) and obtain (also see supplementary)%
\begin{equation}
\frac{R_{s+g}-R_{s}}{R_{s}}=\frac{4}{n_{s}^{2}-1}\frac{4\pi }{c}\sigma
^{\prime }\left( \omega \right) +O\left( \alpha _{\mathrm{QED}}^{2}\right) .
\label{refl}
\end{equation}%
$\sigma ^{\prime }\left( \omega \right) =$Re $\sigma (\omega )$ is the real
part of the complex optical conductivity. %Here  
%$R_{s}$  and $R_{s+g}$ denote the reflection of the SiC substrate (index n$_s$) %and of the substrate with graphene, respectively.  
Eq.(\ref{refl}) leads to two key aspects: (i) the reflection coefficient
provides a direct measurement $\sigma ^{\prime }\left( \omega \right) $ (or
absorption) without reference to any model assumption,e.g., $\Delta R/R =16\pi/({n_{s}^{2}-1})c*\left(\sigma^{\prime}\left(\tau\right)-\sigma^{\prime}\left(0\right)\right)$; (ii) using the
established universal value $\sigma _{0}=\frac{e^{2}}{4\hbar }$ for graphene
monolayer without pulse \cite{RMP2009, KFM2008}, there exists a threshold value for
the photoinduced differential reflectivity that corresponds to the transition to a negative
optical conductivity and thus to gain behavior $\left. \Delta R/R\right\vert _{c}=-\frac{4\pi \alpha _{\mathrm{QED}}}{n_{s}^{2}-1}$. 
%\cite{suppl}.
%\begin{equation}
%\left. \Delta R/R\right\vert _{c}=-\frac{4\pi \alpha _{\mathrm{QED}}}{%
%n_{s}^{2}-1}.
%\end{equation}%
With $n_{s}=2.7$ for SiC substrate follows $\left. \Delta R/R\right\vert
_{c}=-1.4582\%.$, which can be determined experimentally by measuring
differential reflectivity the optical spectra [Fig. 1(d)]. Here the
reflection from the zero conductivity state in the pumped garphene/SiC
sample exactly corresponds to the case of bare SiC substrate.
Critical to note also the negative $\sigma ^{\prime }\left( \omega \right)$ represents the hallmark for the existence of gain in the excited \emph{graphene layer}, which should not be confused with the refection loss in the substrate. 

\begin{figure}[tbp]
\includegraphics[scale=0.33]{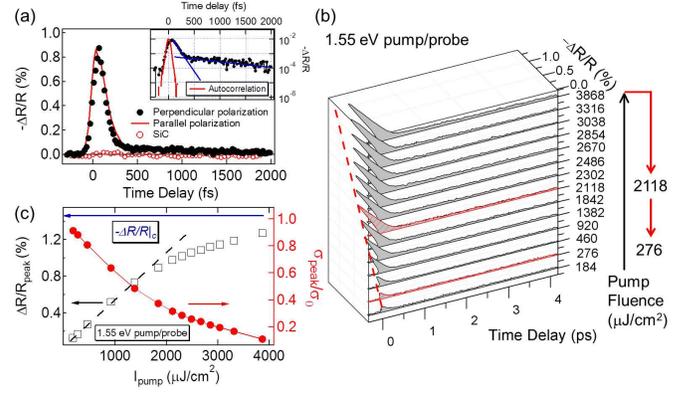}
\caption{(color online) (a) The degenerate differential reflectivity at 1.55
eV for the graphene monolayer (solid dots and line) and SiC substract (empty dots)
under the pump fluence of $1842$ $\protect\mu \mathrm{J/cm}^{2}$. 
 Inset: $%
\Delta R/R$ in logarithmic scale (black dots) together with the pump-probe
autocorrelation (red line). (b), $\Delta R/R$ at 1.55 eV for different pump
fluences. The dashed straight line is guide for the eyes. The different
curves were taken in the order as marked. The perfect overlap of the pair of curves indicates the signals are not
caused by laser-induced permanent changes. (c) The peak ${\frac{\triangle R}{R}}|_{peak}$ as function of the pump
fluence (black squares) and the corresponding conductivity change (red solid
dots). Blue arrow marks the threshold for zero conductivity $\left.
\Delta R/R\right\vert _{c}=-1.4582\%$ (see text). Dashed line: linear
dependence (guide to the eyes).}
\label{Fig2}
\end{figure}

Next we present ultrafast degenerate reflectivity spectroscopy to reveal
femtosecond nonlinear saturation. A typical temporal profile of $\Delta R/R$
at 1.55 eV in the graphene monolayer (black dots) is shown in Fig. 2(a),
clearly showing a negative transient signal. Note the photoinduced change is
negligible in the controlled experiment using the SiC substrate without
graphene (red circles). Several temporal regimes can be identified in the
logarithmic scale plot in the inset: a pulse width limited rise $\sim 40$ 
\textrm{fs} (red line), followed by a bi-exponential decay of $70$ \textrm{fs%
} and $2.5$ \textrm{ps} (blue lines). The power dependence of photoinduced $%
\Delta R/R$ at the maximum of $\Delta R$ in Fig. 2(b) reveals a clear
nonlinear behavior. Fig. 2(c) summarizes $\Delta R/R$ at the signal peak for
different pump fluences $I_{p}$, showing a nonlinear pump fluence dependence
above $\simeq 1850\mu \mathrm{J/cm}^{2}$, at least one order of magnitude
higher than what is reported for semiconducting single-walled carbon
nanotubes \cite{WangPRL2010}. Following Eq.(\ref{refl}), the measured $%
\Delta R/R$ allows us to obtain the corresponding peak conductivity in
photoexcited graphene, as shown by the red dot in Fig. 2(c) (normalized by $%
\sigma _{0}$). At the highest pump fluence used $\simeq 3868\mu \mathrm{J/cm}^{2}$, $\Delta R/R$ peak appoaches 90$\%$ of the critical value $%
\left. \Delta R/R\right\vert _{c}$ and the peak conductivity drops to only 10%
$\%$ of $\sigma _{0}$.

The most striking response is obtained after ultrafast \emph{non-degenerated}
differential reflectivity. Fig. 3(a) shows dynamics using
1.55 eV pump and low energy probes at 1.16 eV and 1.33eV. It is clearly visible that, at high pump fluence, the critical value $\left. \Delta R/R\right\vert _{c}$, the threshold for
negative conductivity (blue lines), indeed occurs for both cases.  
With increasing pump fluence, Fig. 3(b) indicates that, above $I_{pump,c}\simeq
2000\mu \mathrm{J/cm}^{2}$, the stimulated infrared emission
surpasses absorption loss in the photoexcited graphene for the 1.16 eV probe. The temporal
profiles for pump fluence above the gain threshold indicate that the
negative conductivity can persist for hundreds of fs, e.g., the 3960 $\mu \mathrm{J/cm}^{2}$ at 1.16 eV. We emphasize three key
aspects of this conclusion: (i) the critical value has not been reached by
the degenerate pump/probe [Fig. 2(c)] and appears exclusively for
non-degenerate condition when probing below 1.55 eV, (ii) the $\left. \Delta
R/R\right\vert _{c}$ is a \emph{model independent value} that corresponds to 
$\sigma =0$, which directly indicates the transition from loss to gain
behavior, (iii) the femtosecond emergence of stimulated emission even at the 
$\sim$400 meV below the excitation level indicates very rapid establishment
of broadband population inversion and broadband gain in the strongly
photoexcited graphene states. 
%Complementary measurements and analysis at other probe energy are given in the supplementary section \cite{suppl}. 
\begin{figure}[tbp]
\includegraphics[scale=0.4]{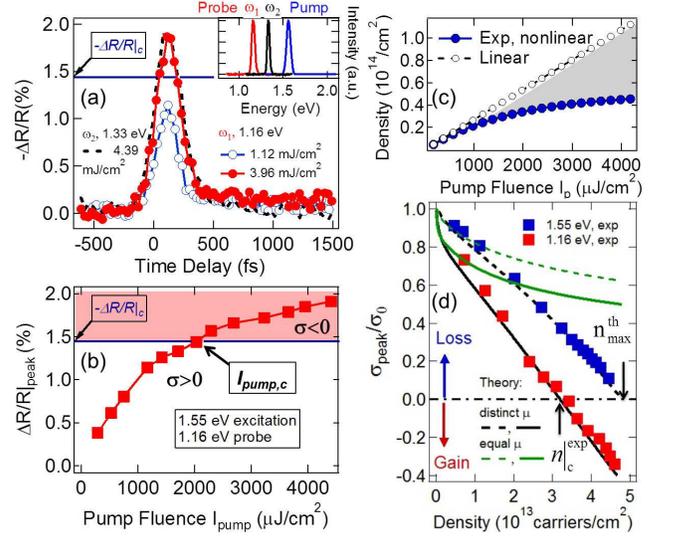}
\caption{(color online) (a) Ultrafast $\Delta R/R$ at 1.55 eV pump, 1.16
eV probe, at 1116 and 3960 $\protect\mu \mathrm{%
J/cm}^{2}$, and 1.33 eV probe, at 4390 $\protect\mu \mathrm{%
J/cm}^{2}$, respectively. Blue arrow marks $\left. \Delta R/R\right\vert _{c}=-1.4582\%$ for zero
conductivity. Shown
together are the pump and probe spectra. (b), The peak transient reflectivity
as function of the pump fluence. 
% clearly shows that the critical value $%
%\left. \Delta R/R\right\vert _{c}$ (blue line) indeed occurs $I_{pump,c}
%\simeq 2000\protect\mu \mathrm{J/cm}^{2}$. 
(c), The extracted transient
fermion density at 40 fs (blue dots), as explained in the text, which is
significantly lower than $A_{0}I_{p}/\hbar \protect\omega $ obtained from
the universal conductivity (open circles), as illustrated in shadow area. (d),
Theory (lines) vs. experimental values (rectangles) for non-degenerate (red)
and degenerate (blue) transient conductivity at 40 fs. Shown together (lines) are two model simulations with the single (green) or distinct (black) chemical 
potentials. }
\label{Fig3}
\end{figure}

We can extract the number of photoexcited electrons in graphene immediately
after the \textrm{\ }laser pulse from 
\begin{equation}
n_{\text{ex}}(I_{p})=\int_{-\infty}^{\infty}\frac{\mathrm{d}t}{\tau_{p}}n_{\text{ex}}(t,I_{p})=\frac{1}{\hbar\omega}\int_{-\infty}^{\infty}\frac{\mathrm{d}t}{\tau_{p}}I(t,I_{p})A\left(t\right),\label{eq: eq2}\end{equation}
%where $I(t)$ is the Gaussian pulse envelope with total pulse fluence $%
%I_{p}=\int_{-\infty }^{\infty }\frac{dt}{\tau _{p}}I(t)$. 
where $I(t,I_{p})$ is the Gaussian pulse envelop $I(t,I_{p})=I_{p}\sqrt{\frac{4\ln2}{\pi}}\exp\left[\frac{-4\ln2}{\tau_{p}^{2}}t^{2}\right]$,
normalized such that the total pulse fluence is $I_{p}=\int_{-\infty}^{\infty}\frac{\mathrm{d}t}{\tau_{p}}I(t,I_{p})$.
For $\tau _{%
\mathrm{th}}\ll \tau _{p}$, the absorption coefficient $A\left( t\right) $
is determined by the adiabatic dependence of the absorption on the pump
fluence and can be derived from the measured saturation curve at 40 fs (Fig.
2c) by $A(t)=A_{0}+\Delta A(t)=A_{0}(1+\frac{\Delta A(t)}{A_{0}})$ (see supplementary). 
% \cite{suppl}. 
Without pump, the absorption of a graphene monolayer on SiC is $A_{0}={%
\frac{4}{(1+n_{SiC})^{2}}} {\frac{\pi e^{2}}{\hbar c}} $. Using the actual
absorption $A\left( t\right) $, instead of $A_{0}$, is crucial as the linear
relation $\hbar \omega n_{\mathrm{ex}}\simeq A_{0}I_{p}$ substantially
overestimates $n_{\mathrm{ex}}$ during the pulse propagation, as shown in
Fig. 3(c). 
%This finding is critical to quantitatively describe the carrier dynamics in strongly photoexcited graphene state %\cite{LuiPRL2010, LuiPRB2010}. 
From Eq.(3), we extract from the data extremely dense photo-excited fermions 
$n_{\mathrm{max}}^{exp }\simeq 0.5\times 10^{14}$ \textrm{cm}$^{-2} $ for
our electron doped sample, surpassing the saturation in a $10$ \textrm{nm} 
\textrm{GaAs} quantum well by more than two orders of magnitude under the
similar excitation condition \cite{chemlaIEEE1984}. 
%This way we obtain the density-dependent gain and absorption in Fig. 3d. 
%, which shows a critical value of  
%$n \vert _{c}^{exp}=0.34\times 10^{14}$ \textrm{cm}$^{-2}$ for the transition to negative conductivity at 1.16 eV.

Next we analyze the transient state at $\tau _{\mathrm{th}}<t<\tau _{\mathrm{%
c}}$. Immediately after the pulse at $\Delta t$ = 40fs, energy of the
electronic system is conserved because no relaxation into the phonon system
has taken place yet. \ In the case of highly excited graphene, the phase
space constraint of the Dirac spectrum leads to an approximate conservation
of numbers of photoexcited holes and electrons, valid to the second order in
the electron-electron Coulomb interaction \cite{Fritz08,Foster09}. A recent
explicit analysis of the short time dynamics by Winzer et al. indicates that
the conservation of hole and electron numbers is a good approximation for
the high excitation regime \cite{WinzerNL}. %Consider Fermi's
%golden rule for electron-electron scattering
%\begin{equation}
%\Gamma _{i\rightarrow f}=\frac{2\pi }{\hbar }\left\vert \left\langle
%f\left\vert V\right\vert i\right\rangle \right\vert ^{2}\delta \left(
%\varepsilon _{s}\left( k\right) +\varepsilon _{s^{\prime }}\left( k^{\prime
%}\right) -\varepsilon _{r}\left( p\right) -\varepsilon _{r^{\prime }}\left(
%p^{\prime }\right) \right) ,
%\end{equation}%
%with $s,s^{\prime },r,r^{\prime }=\pm 1$ referring to the branches of the
%Dirac cone.
%Simultaneous momentum and energy conservation show that the only
%pair scattering processes allowed are those to conserve the number of
%photoexcited holes and electrons individually\cite{Fritz08,Foster09}. 
Consequently, this gives rise to a slow imbalance relaxation and thus to a
population inverted transient state with quasi-conserved occupations of the
two branches of the Dirac cone in our experimental condition. These,
together with the assumption that a decohered, quasi-thermal state leads to
the non-equilibrium transient distribution function ($\varepsilon =vp$): 
\begin{equation}
f_{e\left( h\right) }\left( \varepsilon \right) =\frac{1}{\exp \left( \frac{%
\varepsilon -\mu _{e\left( h\right) }}{k_{B}T_{e}}\right) +1},
\label{distribution}
\end{equation}%
characterized by the electron temperature, $T_{e}$ and two \emph{distinct}
chemical potentials \ $\mu _{e}$ and $\mu _{h}$, for electrons in the upper
and holes in the lower branch of the spectrum, respectively. Note that a
scenario based on a single chemical potential does not explain the
demonstrated population inversion. In thermodynamic equilibrium holds $\mu
_{e}=-\mu _{h}=\mu $ and $T_{e}=T$. In general, $T_{e}$ and $\mu _{e\left(
h\right) }$ are functions of given photon energy, $\hbar \omega $, total
density, and density of photoexcited carriers, $n_{\mathrm{ex}}$ (see
supplementary section). 
The non-equilibrium optical conductivity is calculated in Keldysh
formalism \cite{Keldysh} and follows as 
\begin{equation}
\sigma \left( \omega \right) =\frac{e^{2}}{4\hbar }\left( 1-f_{e}\left(
\hbar \omega /2\right) -f_{h}\left( \hbar \omega /2\right) \right) .
\label{opt}
\end{equation}

Fig. 3(d) compares the peak transient conductivity with the calculated
conductivity $\sigma \left( \omega \right) $ of Eq.\ref{opt} as function of $%
n_{\mathrm{ex}}$ for two probe photon energies 1.55 eV and 1.16 eV.
Excellent agreement between experiment and theory is found which
demonstrates the faithful representation of the transient state at $40$ 
\textrm{fs} by the distribution function, Eq.\ref{distribution}. For the
degenerate scheme, our theory (black dashed line) yields $\sigma \rightarrow
0$ and thus perfect transparency for $n_{max}^{th}=0.48\times 10^{14}$ 
\textrm{cm}$^{-2}$. Once the systems is driven into this regime, a balance
between stimulated emission and absorption will lead to a transprancy. For
non-degenerate scheme by probing at 1.16 eV, our theory (black solid line)
predicts a critical density $n|_{c}^{th}=0.32\times 10^{14}$ \textrm{cm}$%
^{-2}$ for the transition from loss to gain. All of these results agree
quantitatively with the experimental values $n_{max}^{exp}=0.5\times 10^{14}$
\textrm{cm}$^{-2}$ and $n|_{c}^{exp}=0.34\times 10^{14}$ \textrm{cm}$^{-2}$,
respectively (black arrows) \cite{note}. 
%In addition, carrier temperature of the transient state at 40 fs can be derived %from the discussed energy conservation in the electronic system, which is %$\sim$ 2800K at the threshold fluence.   
In addition, the experiment-theory comparison
of the conductivity $\frac{\sigma(\omega)}{\sigma_{0}}$ is shown in Fig. 3d for the distinct- (black lines) and the equal-chemical-potential model (green lines) at the probe photon energy
$\hbar\omega=1.55\text{eV}$ and $1.16\text{eV}$, which clearly identifies the validity of the distinct-$\mu$
model calculation.

\begin{figure}[tbp]
\includegraphics[scale=0.36]{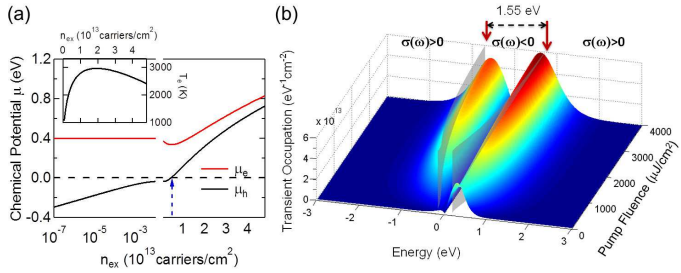}
\caption{(color online) (a) The calculated transient electron ($\protect\mu %
_{e}$), hole ($\protect\mu _{h}$) chemical potentials and transient
temperature (inset) as function of photoexcited carrier density. 
%(b)
%Schematics of a crossover from a hot and dilute Maxwell-Boltzmann gas to a
%dense degenerate Fermi-Dirac system. 
(b) Transient electron and hole
distribution function at 40 fs are plotted for different pump fluence. The
two intersection planes represent the occupation probability of an
electron-hole pair at the given excitation photon energy is equal to unity,
i.e., $f_{e}(\protect\omega/2)+f_{h}(\protect\omega/2)=1$.}
\label{Fig4}
\end{figure}

The detailed information of the transient state at 40 fs as a function of $%
n_{\mathrm{ex}}$ is shown in Fig. 4(a) for the transient chemical potentials
and carrier temperature (inset). For weak excitation $n_{\mathrm{ex}}\ll
\omega ^{2}/v^{2}$ (but larger than the number of thermally excited carriers
before the pulse) we find $k_{B}T_{e}\sim $ $\hbar \omega $ while $\mu
_{h}<0 $, corresponding to a hot, dilute gas of classical holes. Increasing n%
$_{ex}$ changes the sign of the hole chemical potential ($\mu _{h}>0$) and
eventually decreases T$_e$, i.e. as a function of pump fluence we enforce a
crossover from a hot and dilute Maxwell-Boltzmann gas to a degenerate
population inverted quantum system, with $\mu
_{e}+\mu _{h}$ measuring the degree of population inversion \cite{note2}. Most notably,
the transient conductivity, $\sigma \left( \omega \right) $, of Eq.(\ref{opt}%
) changes sign if $\mu _{e}+\mu _{h}$ crosses $\hbar \omega $, with
broadband optical gain due to population inversion for the entire region $%
\omega <\mu _{e}+\mu _{h}$ ($\sigma<0$), as illustrated between two
intersection planes in Fig. 4(c). The separation between the planes is shown
to approach to the pump energy 1.55eV at high excitation density, consistent
with our experiment.

We have showed the existence of pronounced femtosecond
population inversion and near-infrared gain in strongly photoexcited graphene
monolayers. These results clearly reveal the transient electron and hole
potentials are separated on the time scale of 100s of fs. Our
experimental-theory comparison explains well of the absorption saturation
and gain from the dense fermions, and reveals a crossover from a
hot Maxwell-Boltzmann gas to a degenerate dense Fermi-Dirac system.

This work was supported by by the U.S. Department of Energy-Basic Energy
Sciences under Contract No. DE-AC02-07CH11358. 
%We thank A. Patz and I. Chatzakis for their helpful discussions, and X. Huang %and X. Wang for their help with the Raman characterazation of samples. 
J.S. acknowledges support by the Deutsche Forschungsgemeinschaft through the
Center for Functional Nanostructures within subproject B4.5.


\begin{thebibliography}{99}
\bibitem{BonaccorsoNaturephotonics2010} F. Bonaccorso et. al., Nature
Photonics \textbf{4}, 611-622 (2010)
\bibitem{RMP2009} A. H. Castro Neto et al., Rev.
Mod. Phys. \textbf{81}, 109 (2009)
\bibitem{KFM2008} Kin Fai Mak, et al., Phys. Rev. Lett. \textbf{101},196405 (2008).
\bibitem{NairScience2008} R.-R. Nair et al., Science , \textbf{6}, 1308 (2008).
\bibitem{DawlatyAPL2009} J.-M. Dawlaty et al., Appl. Phys. Lett. \textbf{92}, 042116 (2008); P.-A. George et al., Nano Lett. \textbf{8}(12), 4248 (2008); H. Choi et al., Appl. Phys. Lett. \textbf{94}, 172102 (2009).
\bibitem{SunACSNano2010} Z. Sun et al., ACS Nano. \textbf{4}(2), 803
(2010); Q. Bao et al., Advanced Functional Materials \textbf{%
19}, 3077 (2009).
\bibitem{StohrPRB2010} R.-J. St$\ddot o$hr et al., Phys. Rev. B \textbf{82}, 121408 (2010).
\bibitem{LuiPRL2010} C.-H. Liu et al., Phys. Rev. Lett. \textbf{105},
127404 (2010).
\bibitem{LuiPRB2010} W. Lui et al., Phys. Rev. B \textbf{82}, 081408
(2010).
\bibitem{SunNanoLett2010} D. Sun et al., Nano Lett. \textbf{10}(4), 1293 (2010).
\bibitem{BreusingPRL2009} M. Breusing  et al., 
Phys. Rev. Lett. \textbf{102}, 086809 (2009).
\bibitem{NewsonOE2009} R.-W. Newson et al., Optics Express \textbf{17}, 2326 (2009).
\bibitem{SunPRL2008} D. Sun et al., Phys. Rev. Lett. \textbf{101}, 157402 (2008).
\bibitem{oudarPRL1985} J.-L.Oudar et al., Phys. Rev. Lett. \textbf{55}, 2074 (1985)
%\bibitem{note} This value depends on doping, e.g., it is $($0.83)$\cdot %10^{14}$ cm$^{-2}$ for undoped sample. Please refer the supplementary %materials.
\bibitem{GierzNanoLett2008} I. Gierz et al.,  Nano Lett. \textbf{8},
4603 (2010); 
\bibitem{Hupalo09} M. Hupalo et al., Phys. Rev. B \textbf{80}, 041401(R) (2009).


%\bibitem{suppl}  Further methods for data analysis and theoretical modeling are given in the supplementary information.
%, e.g., static and transient optical response, threshold for optical gain, %transient electron/hole densities, optical conductivity for the transient %state, models of two chemical potentials vs. equal potential.
\bibitem{WangPRL2010} J. Wang et al, Phys. Rev. Lett. 
\textbf{104}, 177401 (2010).

\bibitem{chemlaIEEE1984} D. S. Chemla  et al., IEEE J. Quant Electron. \textbf{%
20}, 265 (1984).
\bibitem{Fritz08} L. Fritz et al., Phys. Rev. B \textbf{78}%
, 085416 (2008).

%\bibitem{Muller09} M. M\"{u}ller, J. Schmalian, and L. Fritz, Phys. Rev.
%Lett. \textbf{103}, 025301 (2009).

\bibitem{Foster09} M.\thinspace S. Foster et al., Phys.
Rev. B \textbf{79}, 085415 (2009).

\bibitem{WinzerNL} Torben Winzer et al., Nano. Lett. \textbf{10}, 4839 (2010)

\bibitem{Keldysh}L. V. Keldysh, Sov. Phys. JETP 20, 1018 (1965).
%Zh. Eksp. Teor. Fiz. 47, 1515 (1964) [ 
\bibitem{note} This extremely good agreement between theory and experiment
clearly corroborates $\tau_{th}<40$ fs ($\tau _{\mathrm{th}}<\tau _{\mathrm{p}}<\tau _{%
\mathrm{c}}$). 
%, which places graphene
%in the regime with $\tau _{\mathrm{th}}<\tau _{\mathrm{p}}<\tau _{%
%\mathrm{c}}$. 
The 70 fs and 2.5 ps relaxation components in Fig. 2a can then be naturally
assigned to the cooling of hot carriers $\tau_{\mathrm{c}}$ and
recombination of electron-hole pairs $\tau _{\mathrm{r}}$. 
%Second, his provides extra support for our model with two distinct chemical %potentials at 40 fs in photoexcited graphene (Sect 5.3.2 in suppl. material). 

\bibitem{note2} Our analysis for the low photoexcited carrier density and longer, ps relaxation are consistent with the behaviors found in \cite{kim2011} 

\bibitem{kim2011} R. Kim et al. Phys. Rev. B 84, 075449 (2011)

%\bibitem{note1} In the absence of substrate-induced doping, $n_{\mathrm{ex}%
%}^{\max }$ is 0.57$\cdot 10^{14}$ cm$^{-2}$ (suppl. material).

%\bibitem{BistritzerPRL2009} R. Bistritzer and A. H. MacDonald, \emph{%
%Electronic Cooling in Graphene,} Phys. Rev. Lett. \textbf{102}, 206410 (2009)

%\bibitem{TsePRB2009} Wang-Kong Tse and S. Das Sarma, \emph{Energy relaxation
%of hot Dirac fermions in graphene,} Phys. Rev. B. \textbf{79}, 235406 (2009)

\end{thebibliography}
\end{document}